\documentclass[11pt,twoside]{article}
\usepackage{graphicx}
\usepackage{bm}
\usepackage{subfigure}
\usepackage{multirow}
\usepackage{epstopdf}

\newcommand\nc{\newcommand}
\nc\be{\begin{equation}}
\nc\ee{\end{equation}}
\nc\bea{\begin{eqnarray}}
\nc\eea{\end{eqnarray}}
\nc\beas{\begin{eqnarray*}}
\nc\eeas{\end{eqnarray*}}
\nc{\vect}[1]{\mbox{\boldmath $#1$}}
\nc\lp{\left(}
\nc\rp{\right)}

\begin{document}

\title{Thin films flowing down inverted substrates: Three dimensional flow}
\author{T.-S. Lin and L. Kondic\\
\textit{Department of Mathematical Sciences}\\
\textit{and}\\
\textit{Center for Applied Mathematics and Statistics}\\ 
\textit{New Jersey Institute of Technology, Newark, NJ 07102}\\[1cm]
\textit{A. Filippov}\\
\textit{2920 Shadowbriar Dr., Apt. 134, Houston, TX 77082}}
\date{}

\begin{abstract}
We study contact line induced instabilities for a thin film of fluid under destabilizing gravitational 
force in three dimensional setting. In the previous work (Phys. Fluids, {\bf 22}, 052105 (2010)), we 
considered two dimensional flow, finding formation of surface waves whose properties within the implemented
long wave model depend on a single parameter, $D=(3Ca)^{1/3}\cot\alpha$, where $Ca$ is the capillary 
number and $\alpha$ is the inclination angle. In the present work we consider fully 3D setting and 
discuss the influence of the additional dimension on stability properties of the flow. In particular, 
we concentrate on the coupling between the surface instability and the transverse (fingering) instabilities 
of the film front. We furthermore consider these instabilities in the setting where fluid viscosity varies 
in the transverse direction. It is found that the flow pattern strongly depends on the inclination angle 
and the viscosity gradient. 
\end{abstract}
\maketitle

\section{Introduction}

The problem of spreading of thin films on a solid surface is of interest in a variety of applications, many 
of which were discussed and elaborated upon in excellent review 
articles~\cite{ruschak85,chang,oron_rmp97,stone_04,craster_matar_rmp09}. Perhaps the largest amount of work has 
been done in the direction of analyzing properties of the flow of a uniform film spreading down an incline. 
Starting from a pioneering work by Kapitsa and Kapitsa~\cite{Kapitsa} and progressing to more contemporary 
contributions~\cite{Liu_Gollub93,chang,alekseenko}, a rich mathematical structure of the solutions of governing 
evolution equations, usually obtained under the framework of lubrication (long wave) approach has been uncovered. 
The developed models have lead to evolution equations nowadays known as 
Kuramoto-Sivashinksy~\cite{Chang94,saprykin_pof05}, Benney~\cite{oron_rmp97} and 
Kapitsa-Shkadov~\cite{Trifonov91,chang02}. A variety of nonlinear waves which have been found, as well as 
conditions for their formation were briefly discussed in the Introduction to our earlier work~\cite{lin_pof10}. 
For the purpose of the present work, it is worth emphasizing that the linear and nonlinear waves which were 
discussed in the cited literature resulted from either natural or forced perturbations of the film surface, 
in the setup where inertial effects were relevant: flow of uniform film down an inclined plane (with inclination 
angle $\alpha \le {\pi/2}$) is stable in the limit of zero Reynolds number.

In another direction, there has also been a significant amount of work analyzing different type of instabilities
caused by the presence of fluid fronts, bounded by contact lines where the three phases (gas, liquid, solid) meet.
The fluid fronts are unstable, leading to formation of finger-like structures~\cite{troian_el89} whose properties 
depend on the relative balance of the in-plane and out-of-plane components of gravity~\cite{DK01,KD01}, as well as on 
wetting properties of the fluid~\cite{huppert_82,SD85,deBruyn}. The analysis of the contact-line induced instabilities 
has so far concentrated on the films flowing down an incline, so with $\alpha \le {\pi/2}$. In these configurations, 
fluid surface itself is stable - typically, the only structure visible on the main body of the film involves capillary 
ridge which forms just behind the contact line.

It is also of interest to consider situations where body forces (such as gravity) are destabilizing, as it is the case 
during spreading down an inverted surface, with $\alpha> {\pi/2}$. Such a flow is expected to be unstable, even if 
inertial effects are neglected.   An example of related instabilities are wave and drop structures seen in experimental
studies of a pendant rivulet~\cite{alekseenko96, indeikina97}. Furthermore, if contact lines are present, one may expect 
coupling of different types of instabilities discussed above.   In this context of front/contact line induced instabilities,  
two incompressible viscous fluids in an inclined channel were considered~\cite{segin_jfm05}, while more recently the 
configuration where the top layer is denser than the bottom one was studied~\cite{matar_pof10}. Such configurations were 
found to be unstable, and give rise to large amplitude interfacial waves. 

In addition to mathematical complexity, this setting is of significant technological relevance, in particular in the 
problems where there is also a temperature gradient present, which may lead to a significant variation of viscosity 
of the film. Despite significant progress in study of isoviscous thin film flows, a surprisingly few studies have been 
devoted to analyses of fluid dynamics of thin films with variable viscosity. Meanwhile, such flows are rather common in 
various industrial applications. 


Well-known examples include layers of liquid plastics and paints used for coatings, as well as other materials, whose viscosities 
are strong functions of temperature. Very often, the temperature variations are difficult to detect and prevent. Corresponding 
variations of viscosity affect the processes and, as we will discuss in this work, can be included in the model in the relatively 
straightforward manner.

In our previous work that concentrated on isoviscous flow~\cite{lin_pof10} we found that for a film flowing down an inverted 
plane in 2D, fluid front bounded by contact line (contact point in 2D) played a role of local disturbance that induce 
instabilities. As the inclination angle approaches $180^{\circ}$ (that is, the parameter $D=(3Ca)^{1/3}\cot\alpha$ becomes 
smaller), the fluid front influences strongly the flow behind it and induces waves. In particular, the governing equation, 
obtained under lubrication approximation, is found to allow for three types of solutions. These types can be distinguished 
by the value of the parameter $D$ as follows:\\
$\bullet$ {\it Type 1}: $-1.1\le D$, traveling waves; \\
$\bullet$ {\it Type 2}: $-1.9 \le D < -1.1$, mixed waves; \\
$\bullet $ {\it Type 3}: $-3.0 \le D < -1.9$, waves resembling solitary ones. \\
Solutions could not be found for flows characterized by even smaller values of $D$. We presume that this is due 
to the fact that gravitational force is so strongly destabilizing that detachment is to be expected. It is of 
interest to discuss how the discussion of different wave types extends to three spatial dimensions (3D).

The present paper consists of two related parts with slightly different focus. In the first part, we consider in general terms 
the 3D flow of a film spreading on an inverted surface. The problem is formulated in Sec.~\ref{sec:formu}. In 
Sec.~\ref{sec:riv} we consider instabilities occurring in the flow of a single rivulet, as a simplest example of a 3D flow. 
Fully 3D flow is considered in Sec.~\ref{sec:3d}, where we discuss in particular the interaction between the surface 
instabilities considered previously in the 2D setting~\cite{lin_pof10}, and the transverse fingering instabilities at the front. 
We also briefly comment on the connection of the instability considered here and Rayleigh-Taylor instability mechanism. In the 
second part of the paper, Sec.~\ref{sec:noniso}, we discuss a setting which is perhaps more closely related to applications: 
dynamics of a finite width fluid film characterized by  a nonuniform viscosity which varies in the transverse direction. 
Such a setting may be relevant, for example, to the fluid exposed to a temperature gradient, leading to nonuniform viscosity. 

\section{\label{sec:formu} Problem formulation}

We consider completely wetting fluid flowing down a planar surface enclosing an angle $\alpha$ with horizontal. 
The flow is considered within the framework of lubrication approximation, see e.g.~\cite{lin_pof10}. In particular, 
the spatial and velocity scalings, denoted by $x_c$ and $U$, respectively, are chosen as
\be
x_c=\left(\frac{a^2\,h_c}{\sin\alpha}\right)^{1/3}, \quad U=\frac{\gamma \,h^2_c}{3\mu \,a^2}\,\sin\alpha,
\ee
where $a=\sqrt{\gamma/\rho g}$ is the capillary length, $\gamma$ is the surface tension, $\rho$ is the fluid density, 
$g$ is the gravity, $\mu$ is the viscosity and $h_c\ll x_c$ is the scaling on film thickness. Within this approach, one 
obtains the depth averaged velocity $\vect{v}$
\be
3\mu \vect{v}=\gamma h^2\nabla\nabla^2 h-\rho g h^2\nabla h\cos\alpha+\rho g h^2\sin\alpha\bf{i},
\ee
where 
$h=h(x,y,t)$ is the fluid thickness, $\nabla=(\partial x,\partial y)$, $x,y$ are two spatial variables, $t$ is 
time and $\bf{i}=(1,0)$ is the unit vector pointing in the $x$ direction. Using this expression and the mass 
conservation $h_t + \nabla\cdot(h\vect{v})=0$, we obtain the following dimensionless PDE discussed extensively 
in the literature, see e.g.,~\cite{kondic_siam03,schwartz_pfa89,oron_jp92}
\be
\frac{\partial h}{\partial t} + \nabla\cdot\left[\frac{h^3}{\bar{\mu}}\left( \nabla\nabla^2 h -
D \nabla h + \bf{i}\right)\right] =0.
\label{eq:thin_film_3d}
\ee
The parameter $D=(3Ca)^{1/3}\cot\alpha$ measures the size of the normal gravity, where $Ca=\mu U/\gamma$ is the 
capillary number and $\bar{\mu}$ is the normalized viscosity scaled by the viscosity scale $\mu_0$. To avoid well known 
contact line singularity, we implement precursor film approach, therefore
assuming that the surface is prewetted by a thin film of thickness $b$, see~\cite{lin_pof10} for further discussion
regarding this point.

\section{\label{sec:riv} Inverted rivulet}
 
In this section we consider a single rivulet with front on the underside of a solid surface.   The related problem of an infinite
rivulet was studied in a number of works.  For example, the exact solution of Navier-Stokes 
equation for steady infinite rivulet has been obtained~\cite{tanasijczuk_ejmb10}, and under  
the lubrication assumption the stability of such setting was studied for completely and partially wetting 
fluids~\cite{sullivan_qjmam08,benilov_jfm09}. However, the influence of fluid front on the stability has not been studied.
In the following, we will first extend the steady infinite rivulet solution~\cite{duffy_pof03} to include presence of  a precursor film. 
Then we will examine the effect of fluid front on the stability of an inverted rivulet.

\subsection{Inverted infinite rivulet}
\label{sec:steady}

Consider infinite length rivulet flowing down an inverted planar surface. A steady state shape of the rivulet, independent
of the downstream coordinate $x$ can be found by solving Eq.~(\ref{eq:thin_film_3d}), which in this special case reduces to
\begin{equation}
\left( h^3 h_{yyy} \right)_y -D \left( h^3 h_y \right)_y=0.
\end{equation}
Integrating this equation and applying the boundary conditions $h_y=h_{yyy}=0$ on $\partial \Omega$ and 
integrating again yields
\begin{equation}
h_{yy} -D h = a,
\end{equation}
where $a$ is a constant. Concentrating on inverted case, ie., $D<0$, the general solution can be written as
\begin{equation}
h_r(y) = A\cos(\sqrt{-D} \, y) + B\sin(\sqrt{-D}\, y) -\frac{a}{D}.
\end{equation}
Without loss of generality, we impose the symmetry condition at $y=0$ to find $B=0$, and the complete wetting 
assumption further determines the rivulet's width as 
$\Omega = \left[-\pi / \sqrt{-D}, \pi / \sqrt{-D}\right]$. At the boundaries $h\left(\pm\pi / \sqrt{-D}\right)=b$, 
so we obtain 
\begin{equation}
h_r(y)=\frac{A-b}{2}\cos\left(\sqrt{-D}\, y\right)+\frac{A+b}{2}, 
\label{eq:rivulet_exact}
\end{equation}
where $A=h_r(0)$ is a constant.

According to the above analysis, we find a family of exact rivulet solutions for a given $D$.  The unique solution 
can be obtained by specifying flux or average thickness at the inlet.

\subsection{Inverted rivulet with a front}
\subsubsection{Initial and boundary conditions}

To analyze the effect of the front bounded by contact line (regularized by the precursor) on the rivulet flow, 
we perform numerical simulations of the 3D thin film equation via ADI method. The Appendix gives the complete 
details of the implemented method. The boundary conditions are such that constant flux at the inlet is maintained 
with additional assumption that the shape is taken as a steady rivulet profile. The choice implemented here is 
\begin{equation}
h(0, y, t)=h_r(y), \quad h_{xxx}(0,y,t)-Dh_x(0,y,t)=0.
\end{equation}
In addition, we choose $A=2$ in the steady rivulet solution so that the average thickness at the inlet is $1$. 
At the outlet, $x=L$, as well as at the $y$ boundaries, we assume zero-slope and a precursor film
\begin{eqnarray}
& h(L,y,t)=h(x,\pm M,t)=b,& \\ 
& h_x(L,y,t)=h_y(x,\pm M,t)=0,&
\end{eqnarray}
where $[0,L]$ is the domain size in the $x$ direction, $[-M,M]$ is the domain in the $y$ direction and $b$ is 
typically set to $0.1$.  The initial shape of the rivulet is chosen as a hyperbolic tangent to connect smoothly 
the steady solution and the precursor film at $x=x_f$ - here we choose $x_f=5$. It has been verified that the results 
are independent of the details of this procedure.

\subsubsection{Results}

The presence of the contact line modifies the steady solution discussed in Sec.~\ref{sec:steady}.  Without
going into the details of this modification, for the present purposes it is sufficient to realize that 
the speed of the traveling rivulet, $V_r$, can be easily computed by comparing the net flux with the average 
film thickness as
\begin{equation}
V_r = \frac{\int (h_r^3 - b^3)\, dy}{\int (h_r-b)\, dy } = 
\frac{5}{8}\,A^2 + A\,b + O(b^2).
\end{equation}
As we will see in what follows, $V_r$ is important for the purpose of understanding the computational results
in the context of 2D instabilities discussed previously~\cite{lin_pof10}.  

\begin{figure}[htb]
\centering
\includegraphics[scale=0.4]{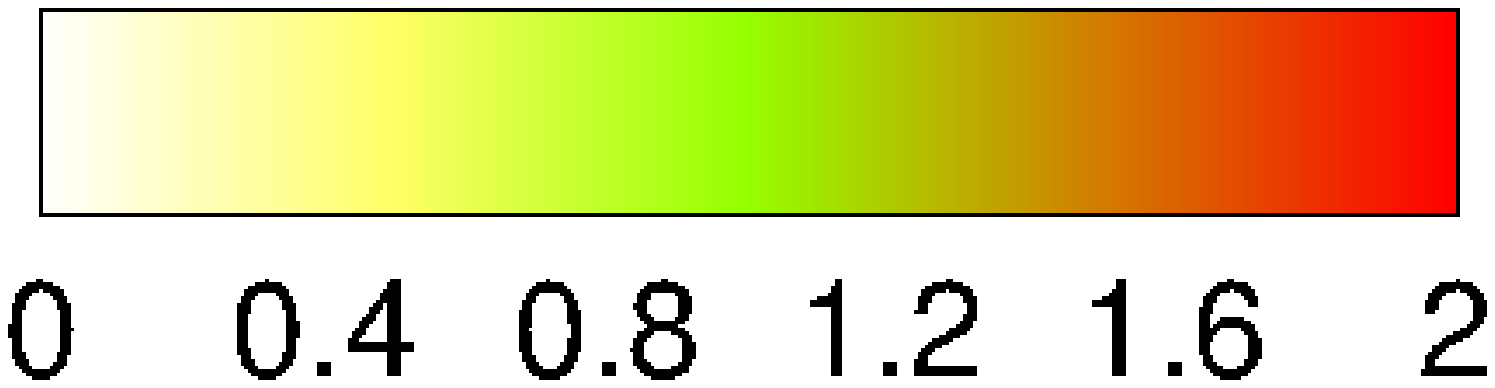}
\includegraphics[scale=0.5, angle=-90]{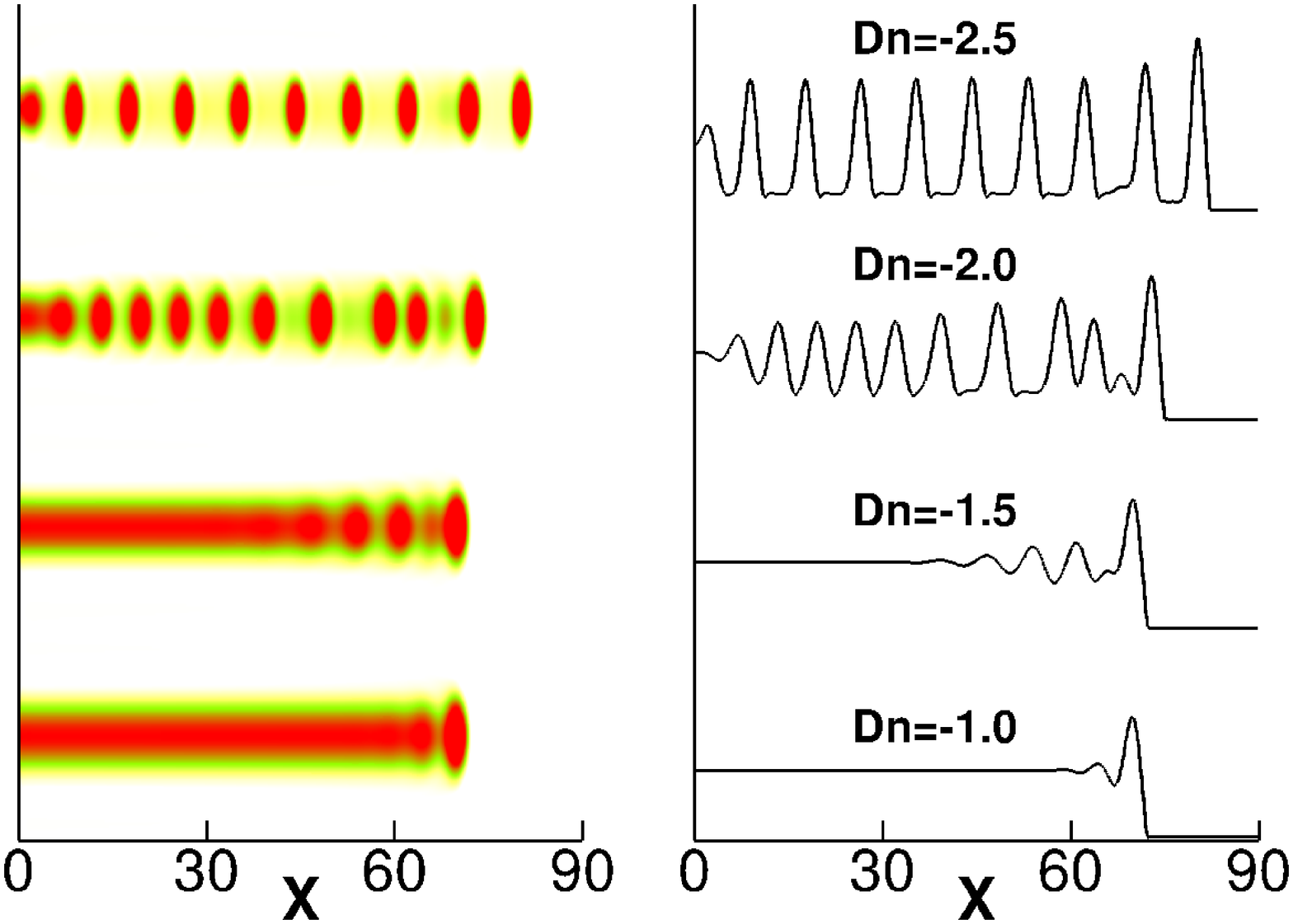}
\caption{(Color online) Rivulet flow for different $Dn$'s at $t=25$.  The domain size is specified by $L=90$ and $M=5$.
Left hand side shows the contour plot and the right hand side shows the cross section ($y = \mbox{const}$.) at 
the middle of a rivulet. We will use similar way of presenting results, and the same color map in the other 
figures given in this paper.}
\label{fig:rivulet}
\end{figure}

Figure~\ref{fig:rivulet} shows the computational results for different $D$'s at $t=25$.  In order 
to compare with 2D simulations, we have renormalized $D$ with 
$V_r$ (recall that the 2D traveling wave velocity, $V$, equals $1+b+b^2$  while $V_r=2.7$ for $A=2$ in our 3D rivulet 
simulations). We call the renormalized $D$ by $Dn = D/(V_r)^{1/3}$.

For $Dn=-1.0$, we still observe traveling type solution. We have examined speed of the capillary ridge 
and found that it equals $V_r$, as predicted. For $Dn$ larger in absolute value, the traveling type solution becomes 
unstable and waves keep forming right behind the capillary ridge. For $Dn > -1.5$, simulations suggest that 
the instability is convective, since it is carried by the flow and moves downstream from the initial  
contact line position, $x_f$. For $Dn < -1.5$, the instability is absolute. At the time shown, the whole rivulet is 
covered by waves which in the cross section resemble solitary ones.

The rivulet simulations show a qualitative similarity to our 2D simulations (vis. the right hand side of 
Fig.~\ref{fig:rivulet} and Fig.~4 in~\cite{lin_pof10}). Therefore, on one hand this result validates the accuracy 
of our 3D simulations. On the other hand, it also suggests that the instability regimes, ({\it type 1, 2} 
and {\it 3}) can be extended to 3D rivulet geometry. Clearly, it was necessary to use renormalized value of $D$, 
$Dn$, to be able to carry out this comparison.

It is also of interest to relate the present results to stability properties of an inverted infinite length rivulet without
a front.   In this case, it is known that there exists a critical angle between $\pi/2$ and $\pi$ such that the inverted 
infinite rivulet is unstable if the inclination angle is larger than the critical one~\cite{benilov_jfm09}.    In order to
be able to directly compare the two problems (with and without a front),  we have carried out simulations of an inverted infinite 
rivulet.  The steady state is fixed  by choosing $A=2$.   We perturb the rivulet at $t=0$ by a single perturbation defined by 
$h(x,y,0) = h_r(y) (1 + A_r sech(x-x_p))$ with $A_r = 0.2$ and $x_p = 5.0$.  In this work we consider only this type of 
perturbation and do not discuss in more detail the influence of its properties on the results.   For this case, we 
find that an inverted infinite rivulet is 
unstable for $Dn\le -0.74$, which is consistent with the fact that there exists a critical angle for stability~\cite{benilov_jfm09}.    An obvious
question to ask is which instability is dominant for sufficiently small $Dn$'s, such that both front-induced and surface-perturbation
induced instabilities are present.   This question will also appear later in the context of thin film flow - to avoid repetition we consider
it for that problem, in Sec.~\ref{sec:rt}.

\section{\label{sec:3d}Inverted film with a front}

We proceed with analyzing stability of an inverted film with a front flowing down a plane. In Sec.~\ref{sec:lsa} we 
extend the results of the  linear stability analysis in the transverse direction to the inverted case. Then, we proceed 
with fully nonlinear time dependent simulations in Sec.~\ref{sec:sim}.  We start with addressing a simple case where 
we perturb the fluid front by a single wavelength only -- this case allows us to correlate the results with the 2D surface 
instabilities discussed previously~\cite{lin_pof10}, with the instabilities of a single inverted rivulet discussed in 
Sec.~\ref{sec:riv}, and also with the well known results for transverse instability of a film front flowing down an inclined 
plane, see e.g.~\cite{DK01}. We proceed with more realistic simulations of a front perturbed by a number of modes with random 
amplitudes, where all discussed instability mechanisms come into play. To illustrate complex instability evolution, we also 
include animations of the flow dynamics for few selected cases~\cite{movie_SM}. We conclude the section by discussing in 
Sec.~\ref{sec:rt} the connection between the instabilities considered here with Rayleigh-Taylor type of instability of an 
infinite film flowing down an inverted plane.

\subsection{Linear stability analysis in the transverse direction}
\label{sec:lsa}

In order to analyze the stability of the flow in the transverse, $y$, direction, we perform a linear 
stability analysis (LSA). The results of similar analysis were reported in previous works, see, 
e.g.,~\cite{kondic_siam03}, but they typically concentrated on the downhill flows, with $D\ge 0$. Here we 
extend the analysis to also consider films on an inverted surface, with $D<0$.  

Consider a moving frame defined by $s=x-Vt$, and assume a solution of the form 
\begin{equation}
h(s,y,t)=H(s)+\epsilon h_1(s,y,t),
\end{equation}
where $\epsilon \ll 1$, $H(s)$ is the traveling wave solution with the speed $V$. 
Then, plug this ansatz into Eq.~(\ref{eq:thin_film_3d}). The leading order term 
($O(\epsilon^0)$) gives the 2D equation
\begin{equation}
-VH'+[H^3(H'''-DH'+1)]'=0
\label{eq:traveling_wave}
\end{equation}
while the first order term ($O(\epsilon^1)$) yields
\begin{eqnarray}
\frac{\partial h_1}{\partial t} &=& 
-\nabla\cdot\left[H^3\nabla\nabla^2h_1 + 3H^2h_1\nabla\nabla^2H\right] \nonumber\\
& &+D\nabla\cdot\left[H^3\nabla h_1+3H^2h_1\nabla H\right]\nonumber\\
& &-\left(3H^2h_1\right)_{s}+Vh_{1s},\label{eq:LSA1}
\end{eqnarray}
where $\nabla=(\partial_{s}, \partial_y)$. The next step is to express the solution, $h_1$, as a 
continuous superposition of Fourier modes, 
\begin{equation}
h_1(s, y, t) =\int^0_{-\infty} \phi(s,q) e^{\sigma t} e^{i q y} dq,
\label{eq:h_1_q}
\end{equation}
where $q$ is the wavenumber and $\sigma$ is the growth rate that determines the temporal evolution 
of $h_1$. For a given $q$, there is an associated eigenvalue problem, see e..g,~\cite{kondic_siam03}. 
The largest eigenvalue corresponds to the growth rate, which is the quantity of interest.

\begin{figure}[htb]
\centering
\includegraphics[scale=0.3, angle=-90]{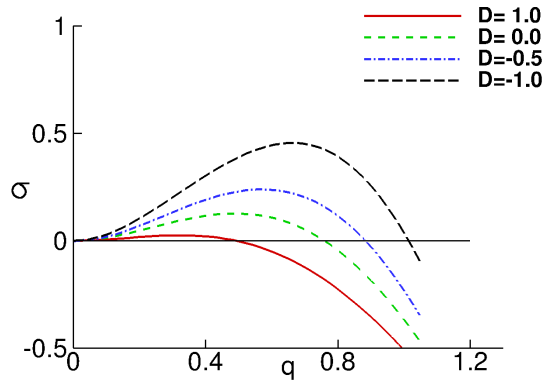}
\caption{(Color online) Wavenumber, $q$, and corresponding growth rate $\sigma$ for different $D$'s.}
\label{fig:HF_LSA}
\end{figure}

Figure~\ref{fig:HF_LSA} shows the LSA results. Each curve represents the corresponding largest eigenvalue for a given 
wavenumber and for fixed $D$. One can see that sufficiently long wavelengths are unstable. Consequently, there is a 
critical wavenumber, $q_c (D)$, which determines the range of unstable wavenumbers to be $[0, q_c]$.  Concentrating now on 
negative $D$'s, we find that $q_c$ increases with $|D|$, suggesting shorter unstable wavelength for larger $|D|$'s, which 
are furthermore expected to grow faster. Therefore, for $D<0$, as $|D|$ is increased (for example by increasing the angle 
$\alpha$, or by making the film thicker), LSA predicts formation of more unstable fingers spaced more densely.

One may note that Fig.~\ref{fig:HF_LSA} only shows the results for $D$ down to $-1.0$. The reason is that a 
base state could be found only for $D\ge -1.1$, and therefore LSA could not be carried out for smaller $D$'s. 
This issue was discussed in~\cite{lin_pof10}, where we were able to find traveling wave solutions only for 
$D$'s in {\it type 1} regime, and could not find such solutions in {\it type 2} and {\it 3} regimes.

\subsection{Fully 3D simulations}
\label{sec:sim}

In this section we discuss the results of fully 3D simulations of the thin film equation, 
Eq.~(\ref{eq:thin_film_3d}). Throughout this section as our initial condition we chose the same
profile as in 2D simulations~\cite{lin_pof10} - that is, two flat regions of thickness $h=1$ and $h=b$
connected at $x=x_f$ by a smooth transition zone described by a hyperbolic tangent,  perturbed
as follows 
\[
x_f(y) = x_{f0} - A_0 \cos(2 \pi y/ \lambda),
\]
where $\lambda$ is the wavelength of the perturbation and $x_{f0}$ is the unperturbed position. Here we 
choose $x_{f0}=5$. The boundary conditions in the flow direction are such that constant flux at the inlet 
is maintained, while at $x=L$, we assume that the film thickness is equal to the precursor. The boundary 
conditions implemented here are
\begin{eqnarray*}
& h(0,y,t)=1, \quad h_{xxx}(0,y,t)-D\,h_x(0,y,t)=0,& \\
& h(L,y,t)=b, \quad h_x(L,y,t)=0.&
\end{eqnarray*}
For the $y$ boundaries, we use the no-flow boundary conditions as
\begin{eqnarray}
h_y(x,0,t)=h_y(x,M,t)=0, \label{eq:full3d_by}\\
h_{yyy}(x,0,t)=h_{yyy}(x,M,t)=0,\nonumber
\end{eqnarray}
where $M$ is the width of the domain in the $y$ direction.

\begin{figure*}[thb]
\centering
\includegraphics[scale=0.6, angle=-90]{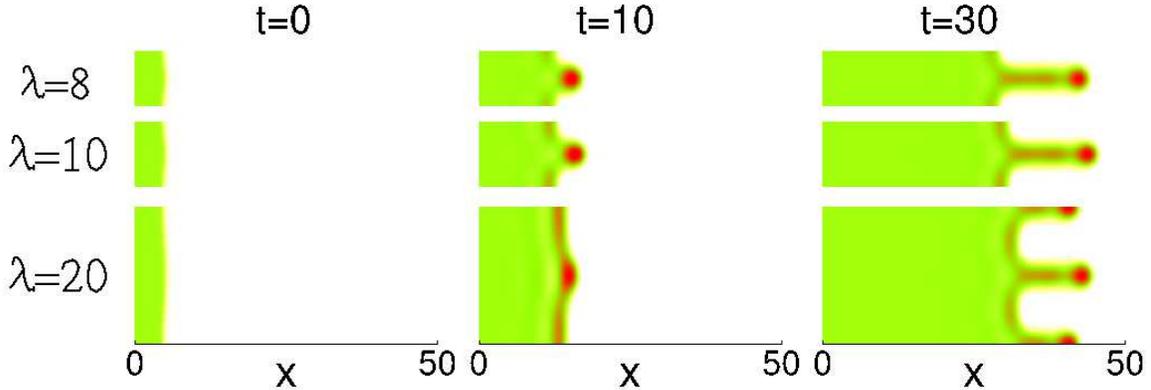}
\caption{(Color online) Time evolution for perturbations of different wavelengths ($\lambda$) and  $D=-1.0$. 
Here the domain is specified by $L=50$ and $M=\lambda$.
}
\label{fig:HF1}
\end{figure*}

Figure~\ref{fig:HF1} shows the results of simulations for $D=-1.0$ perturbed by specified single mode perturbations. For 
$\lambda=8$ and $\lambda=10$, the perturbation evolves into a single finger, whereas for $\lambda=20$, which corresponds 
to the wavelength larger than the most unstable one (vis. Fig.~\ref{fig:HF_LSA}), we observe a secondary instability: 
in addition to the finger that corresponds to the initial perturbation, there is another one (which appears as half 
fingers at $y=0, M$), developing at later time (in this case after $t=10$). This phenomenon can be explained by the 
fact that the domain is large enough to allow for two fingers to coexist. 

\begin{figure}[htb]
\centering
\includegraphics[scale=0.3, angle=-90]{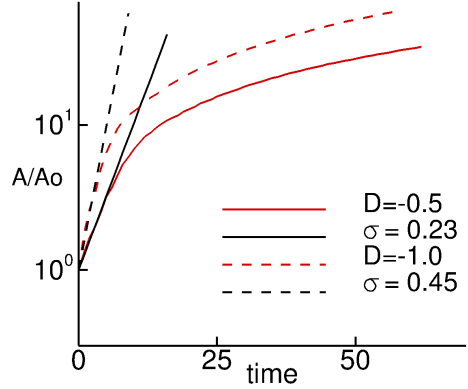}
\caption{(Color online) Length of a finger, $A$, divided by the initial length, $A_0$.  Here $\lambda=M=10$. Red (curved) lines 
show the computational results while the black (straight) lines show the LSA prediction for two different $D$'s.}
\label{fig:HF_FRONT}
\end{figure}

Next, we compare the growth rate of a finger from 3D simulation with the LSA results. This comparison is shown in 
Fig.~\ref{fig:HF_FRONT} for $D=-0.5$ and $D=-1.0$. The initial condition for these simulations is chosen as a single mode 
perturbation, $\lambda=M=10$. The growth rate is extracted by considering the finger's length, $A$, defined as the distance 
between tip and root. As shown in Fig.~\ref{fig:HF_FRONT}, for early times, the finger grows exponentially with the same 
growth rate as predicted by the LSA. For later times, the finger length exhibits a linear growth. This late time behavior 
can be simply explained by the fact that the finger evolves into the rivulet solution discussed in Sec.~\ref{sec:riv}.

\begin{figure*}[thb]
\centering
\includegraphics[scale=0.6, angle=-90]{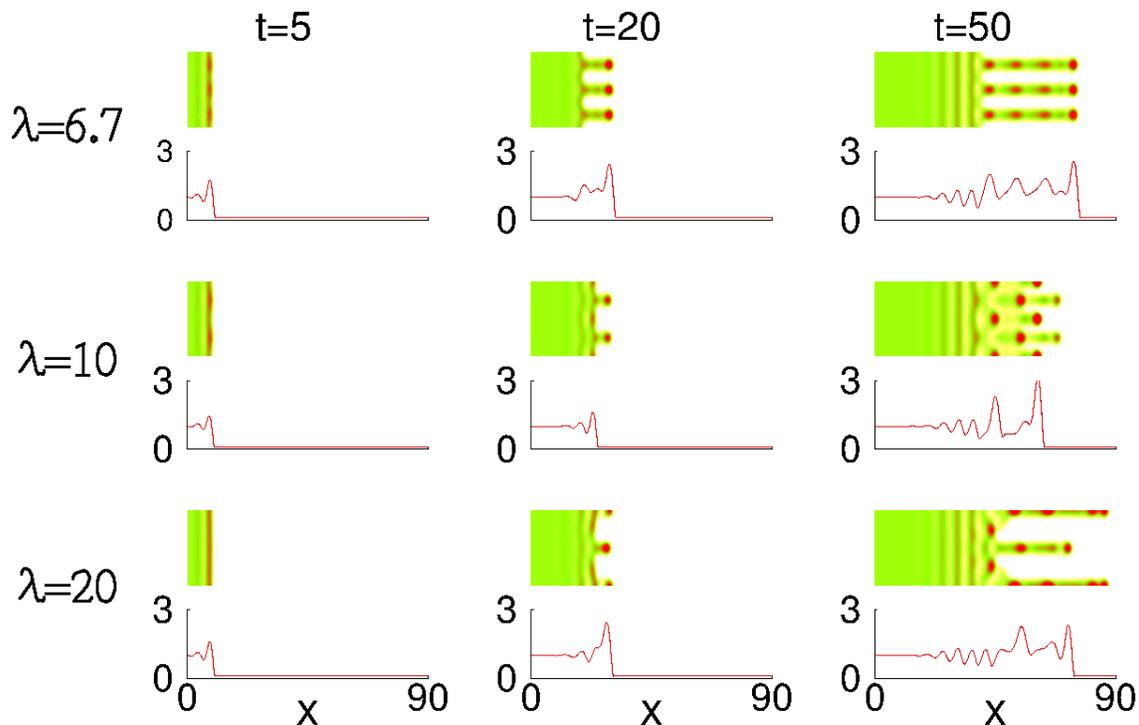}
\caption{(Color online) Time evolution  for perturbations of different wavelengths ($\lambda$) at $D=-1.5$. The domain 
is chosen as $L=90$ and $M=20$. In each sub-block, the upper figure shows the contour plot, and the lower figure shows 
the cross section at $y=10$.}
\label{fig:HF2}
\end{figure*}

Figure~\ref{fig:HF2} shows the numerical results for $D=-1.5$ with three single mode perturbations in a fixed domain, 
$[0,90]\times [0,20]$. Both contour plots and the cross sections at middle of the domain ($y=10$) are presented. In this 
case, in addition to the secondary instability in the case of the perturbation by $\lambda = 20$, we also see secondary 
instability developing for $\lambda = 10$, suggesting that as the absolute value of (negative) $D$ grows, shorter and 
shorter wavelengths become unstable. For this $D$, even $\lambda = 5$ is unstable. This result is consistent with the 
trend of the LSA results shown in Fig.~\ref{fig:HF_LSA}; note however that LSA cannot be carried out for such small $D$'s 
due to the fact that traveling wave solution could not be found. Instead of traveling waves, we find surface instabilities 
on the film. As shown in the contour plot at $t=50$, several red(dark) dots, which represent waves, appear on the fingers, 
as it can be seen in the cross section plots as well. The red(dark) dots keep forming, moving forward and interact with 
the capillary ridge in the fronts. This interaction can be seen much more clearly in the animations available as supplementary 
materials~\cite{movie_SM}.

The cross section plots in Fig.~\ref{fig:HF2} show formation of solitary-like waves. Behind these solitary-like waves, 
there exists a second region where waves appear as `stripes' (vis. the straight stripes in the middle part of the contour 
plot at $t=50$). These stripe-waves move forward for a short distance and then break into several waves localized on the 
surface of the finger-like rivulets. In the cross-section plots, these strip-like waves appear as sinusoidal waves. 
Finally, flat film is observed in the region far behind the contact line. The appearance of such a flat film indicates 
that the flow instability is of convective type. Therefore, for this $D$, the contact line induced waves are carried by 
the fluid and they eventually move away from any fixed position.

\begin{figure*}[htb]
\centering
\includegraphics[scale=0.6, angle=-90]{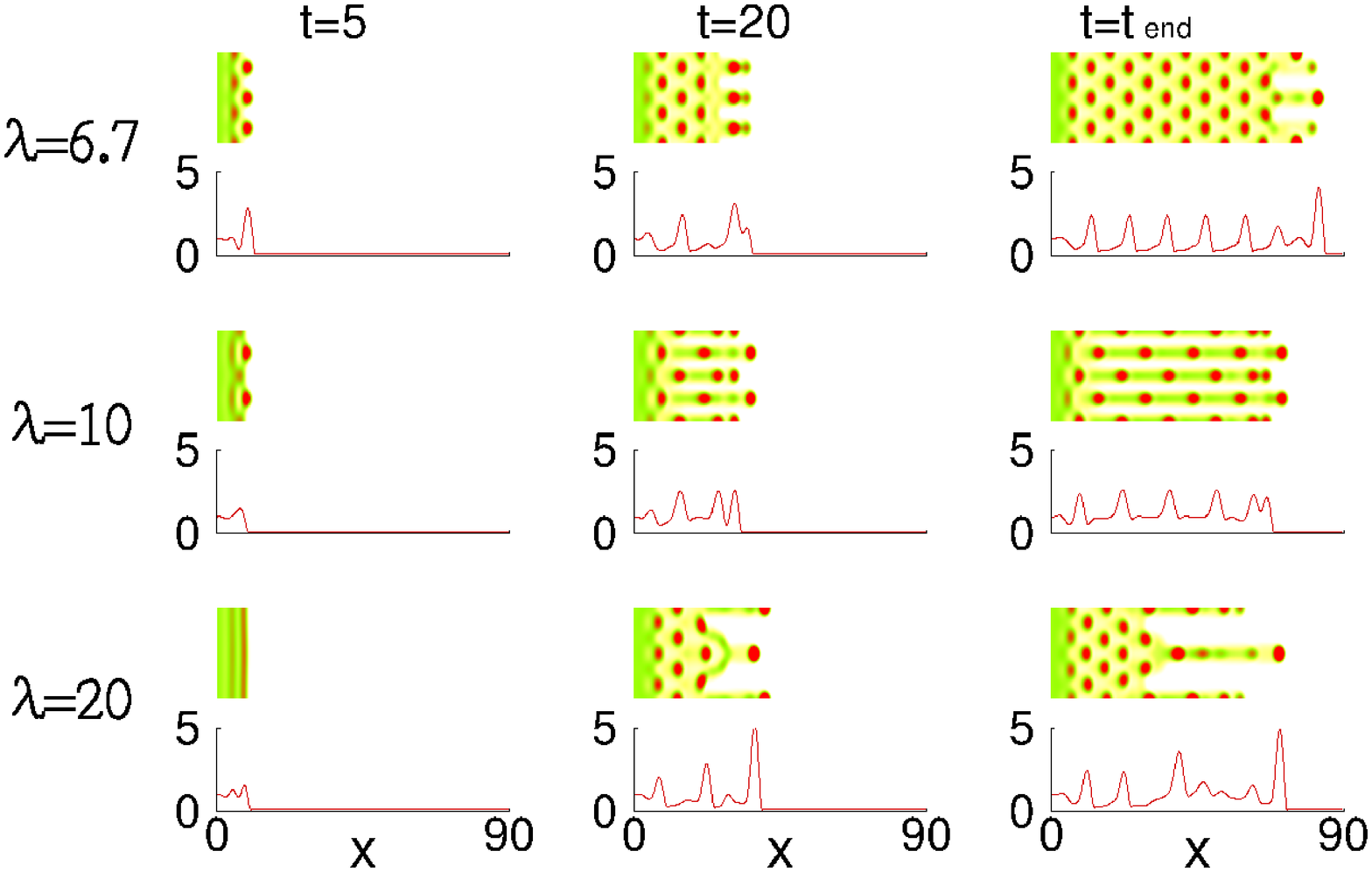}
\caption{(Color online) Time evolution for perturbations of different wavelengths ($\lambda$) at 
$D=-2.0$. $t_{end}=50$ for $\lambda=6.67$, $t_{end}=40$ for $\lambda=10$, $t_{end}=30$ for $\lambda=20$. The domain 
is chosen as $L=90$ and $M=20$. In each sub-block, the upper figure shows the contour plot, and the lower figure 
shows the cross section at $y=10$.}
\label{fig:HF3}
\end{figure*}

Figure~\ref{fig:HF3} shows the numerical results for $D=-2.0$ with the same set of initial perturbation as in 
Fig.~\ref{fig:HF2}. Both contour plot and the cross sections at the middle of the $y$ domain are presented. As 
mentioned in~\cite{lin_pof10}, $D=-2.0$ corresponds to the {\it Type 2} regime and is of absolute instability 
type. This is exactly what we see in the contour plot.   Localized waves shown as red (dark) dots appear all over the 
surface and  we do not see neither strip waves nor flat film appearing. In the cross section plots, we only see 
solitary-like waves.

\begin{figure*}[htb]
\centering
\includegraphics[scale=0.6, angle=-90]{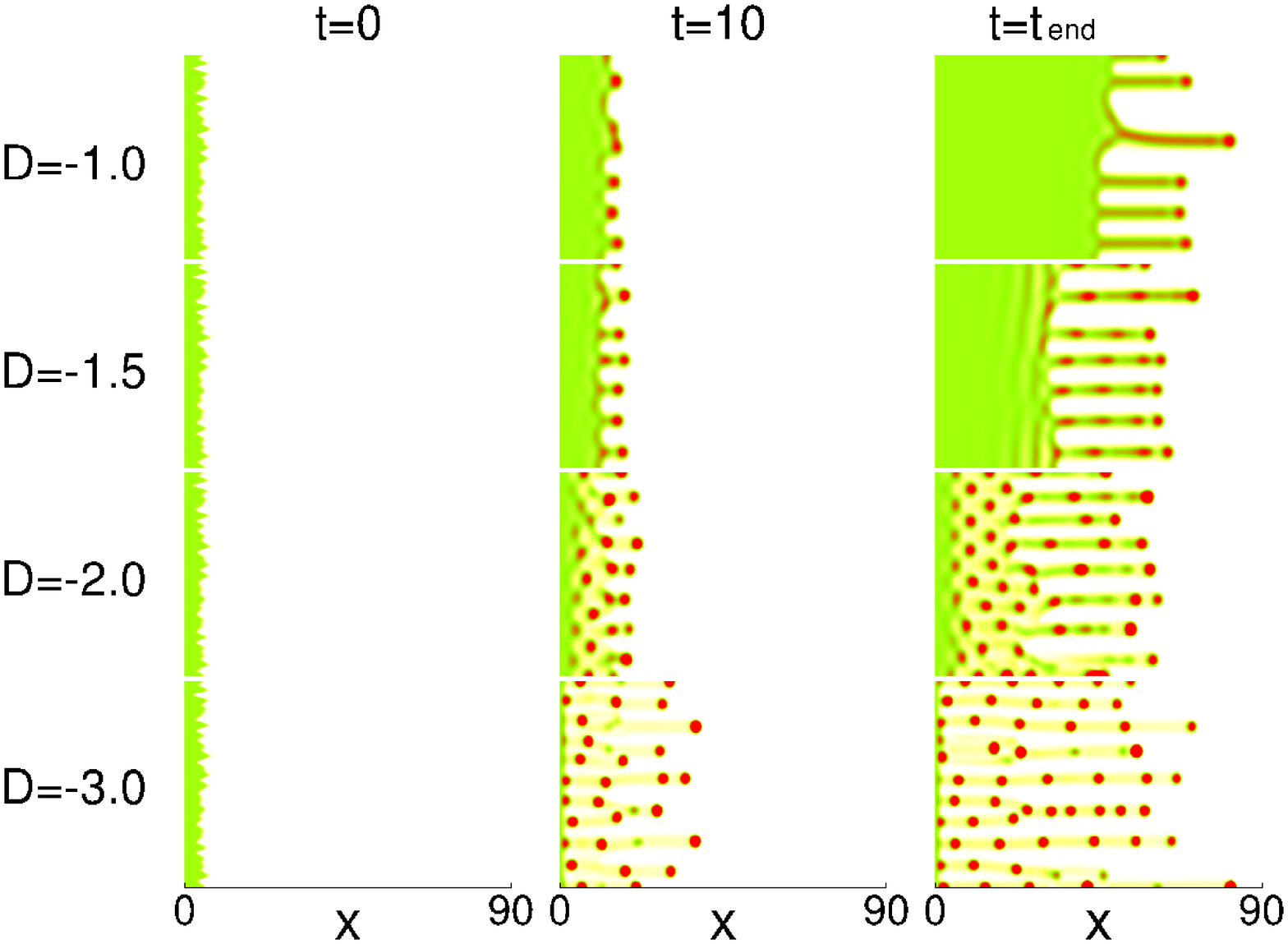}
\caption{(Color online) Time evolution of perturbations for different $D$'s with the same random initial 
perturbations.  Here $t_{end}=50$ for $D=-1.0$, $t_{end}=40$ for $D=-1.5$, $t_{end}=30$ for $D=-2.0$, 
$t_{end}=20$ for $D=-3.0$. The domain is chosen as $L=90$ and $M=50$ See also the animations 
available~\cite{movie_SM}.}
\label{fig:HF4}
\end{figure*}

Next we proceed to analyze the behavior in the case where initially multiple perturbations are present. 
The imposed perturbation consists of $50$ sinusoidal modes with amplitudes randomly selected from $[-0.2, 0.2]$
\[
x_f(y) = x_{f0} - \sum_{i=1}^{50}\,A_i\,\cos((i-1) \, \pi \, y), \quad 
-0.2 \leq A_i \leq 0.2 \, .
\]
Figure~\ref{fig:HF4} shows the simulations for different $D$'s with the same random initial perturbations. The 
initial profile is shown in Fig.~\ref{fig:HF4} ($t=0$). The initial condition is set to be the same for 
all $D$'s so that the non-dimensional parameter $D$ is the only difference  between the 4 panels in Fig.~\ref{fig:HF4}. 

The first obvious observation is that the number of fingers increases as the absolute value of (negative) $D$ becomes larger. 
This is due to the fact that the most unstable wavelength decreases as $|D|$ of (negative) $D$ becomes larger. Therefore more 
fingers can fit into the flow domain. Secondly, the fingers become more narrow for these $D$'s, consistently with the 
predictions for rivulet flow. Thirdly, the absolute/convective instability argument in~\cite{lin_pof10} is a good explanation 
for these contact line induced instabilities: there is no surface instability seen for $D=-1.0$; we see convective instability, 
shown as localized waves/stripes/flat film for $D=-1.5$; and absolute instability for $D=-2.0$ and $D=-3.0$. 

{\bf Remark}

Here we comment on two additional sets of simulations - corresponding figures are omitted for brevity. One set 
involves the case when the initial condition is chosen as $y$-independent. Mathematically such initial condition 
reduces the problem to 2D, and the solution should remain $y$-independent for all times. However, this is not the 
case in numerical simulations. Numerical errors are present and grow with time. To estimate this effect, one can 
calculate the largest grows rate in the $x$ and $y$ direction based on the LSA results, and estimate the time for 
which the numerical noise becomes significant. For example, for $D=-1.0$, the largest growth rate for instability 
of a flat film in the $x$ direction is $0.25$ and the largest growth rate in the $y$ direction is $0.46$. 
That is, it takes $70$ time units for noise of initial amplitude $10^{-16}$ (typical for double precision computer 
arithmetic) to grow to $10^{-2}$. So as long as the final time is less than $70$, the numerical noise is still not 
visible. By carrying out such an analysis, we are able to distinguish between the numerical noise induced instability 
and the contact line induced one, and further separate the effect of numerical noise.

The other set or simulations has to do with the case when the initial single mode perturbation is chosen as a stable one. 
In such a case, the amplitude of perturbation decays exponentially and the surface profile soon becomes $y$-independent. 
Again, after sufficiently long time, numerical noise will become relevant and break the $y$-independence.

\subsubsection {The width of a finger}

It is of interest to discuss how fingers' width depend on $D$.  As a reminder, the LSA shows that there exists a most unstable 
wavenumber, $q_m(D)$, and the distance between two neighboring fingers in physical experiments, in the presence of natural or 
other noise, is expected to center around the most unstable wavelength, $\lambda_m = 2\pi/q_m(D)$. On the other hand, 
the width of the rivulet part of a single finger is not known  to the best of our knowledge.  In the following we 
define this width and discuss how it relates to the LSA results.

\begin{figure}[htb]
\centering
\subfigure[$D = -0.5$.]{\includegraphics[scale=0.3, angle=-90]{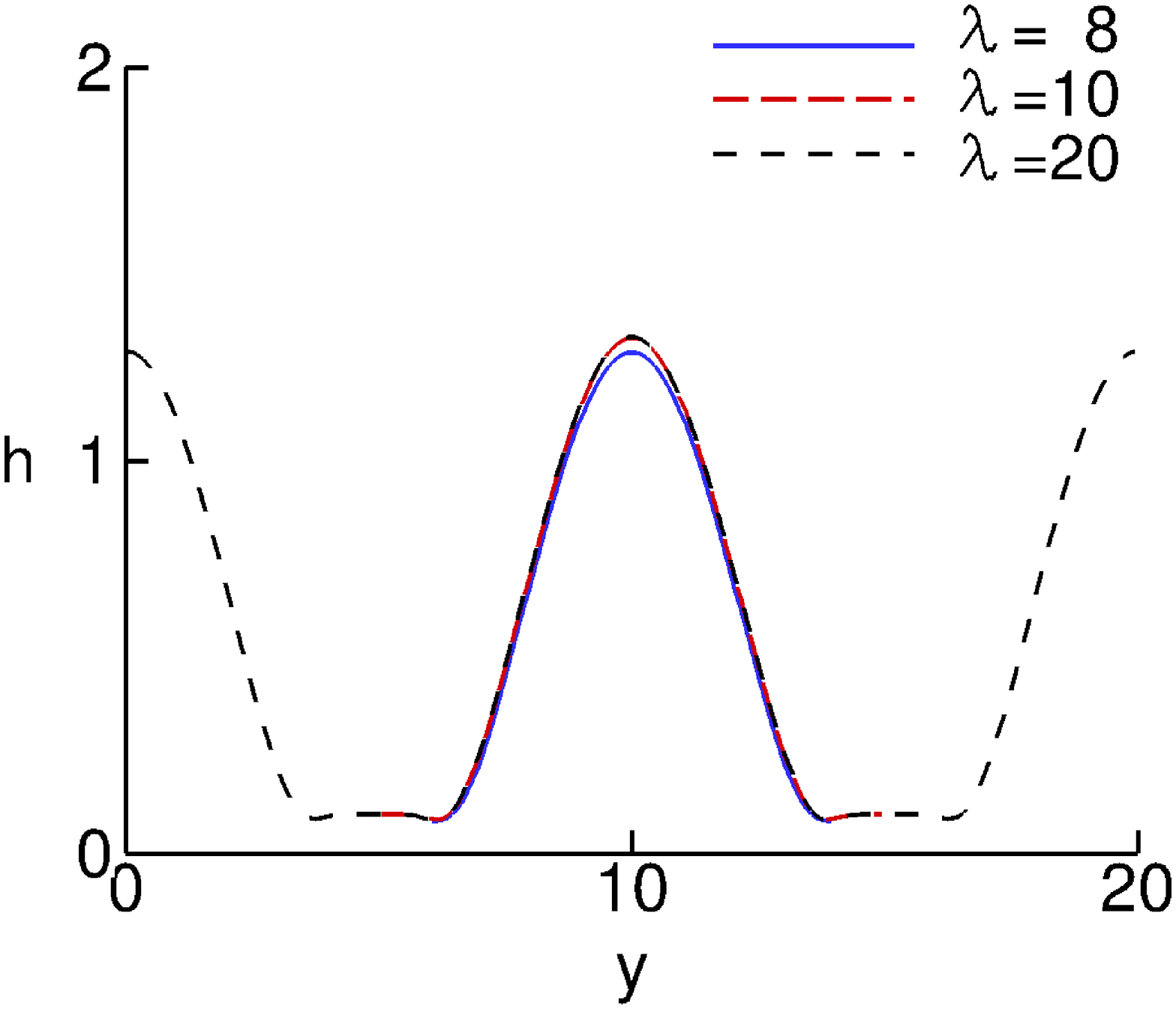}}\\
\subfigure[$D=-1.0$.]{\includegraphics[scale=0.3, angle=-90]{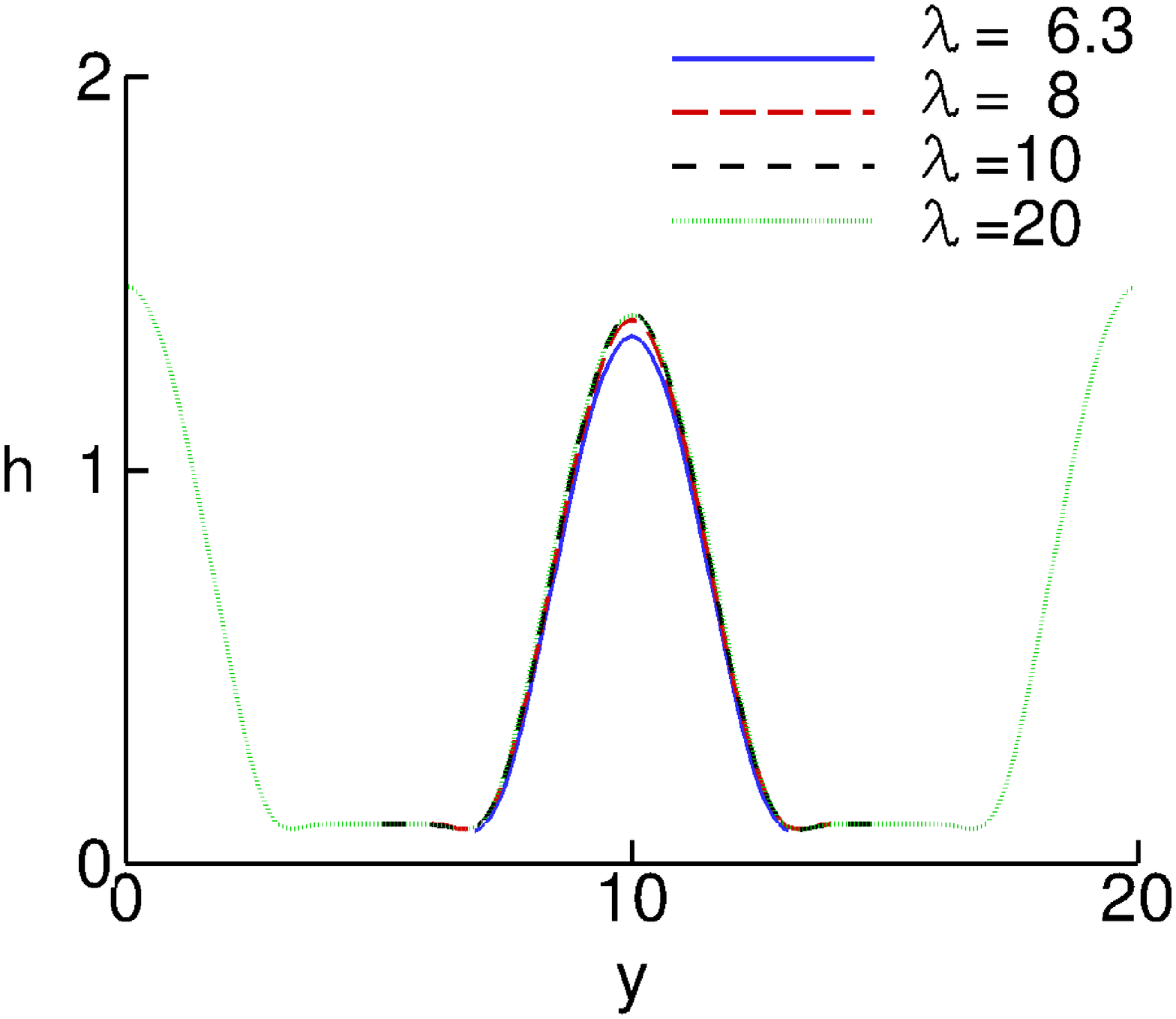}}
\caption{(Color online) 
Cross sections of film thickness as a function of the transverse coordinate, $y$,  for different wavelengths of
initial perturbation, $\lambda$.   
The cross sections are taken from the rivulet part of a finger,  at $x=35$, $t=30$; some of results from which 
the cross sections are extracted can be seen in Fig.~\ref{fig:HF1}. 
The centers of the cross sections are shifted to $y=10$ for the purpose of comparison.   For $D=-0.5$, $\lambda = 6.3$ 
is stable (not shown).
}
\label{fig:HF_slice}
\end{figure}

First we check whether there is a difference between the fingers for different single mode perturbations. 
Figure~\ref{fig:HF_slice} shows the $y$-orientation cross section of the fingers' rivulet part for $D=-0.5$ and $D=-1.0$. 
As shown in the figure, for a given $D$, the fingers are very similar in the cross section, with their shape almost independent 
of the initial perturbation. Note that this result still holds even for the $\lambda$'s that are very close to the critical 
one, $\lambda_c = {2 \pi/q_c(D)}$ - eg., see $\lambda=6.3$ in Fig.~\ref{fig:HF_slice}(b) (here $\lambda_c \approx 6.25$). By 
direct comparison of the parts (a) and (b) of this figure, we immediately observe that the fingers are more narrow for more
negative $D$'s.

To make this discussion more precise, we define the width of a rivulet, $w$, as the distance between two dips on each side of 
a finger (the two dips are shown at  $y\approx 7$ and $y\approx 13$ in Fig.~\ref{fig:HF_slice}).  {\it The main finding is 
that} $w \approx \lambda_c$. This finding applies for all $D$'s and all perturbation wavelengths that we considered. In 
particular note that Fig.~\ref{fig:HF_slice} shows that $w$ becomes smaller as absolute value of (negative) $D$ increases, consistently 
with the decrease of $\lambda_c$, see Fig~\ref{fig:HF_LSA}.

While it is clear that $w$ can not be larger than $\lambda_c$  (since $w$ is independent of the initial perturbation, and for 
$\lambda>\lambda_c$ it must be that $\lambda> w$), at this point we do not have a precise argument why $w$ is so close to 
$\lambda_c$ for all considered perturbations and the values of $D$.   Of course, one could argue that $w\approx \lambda_c$ is 
also consistent with stability of any perturbation characterized by $\lambda < \lambda_c$, since such a perturbation cannot 
support a finger of $w\approx \lambda_c$.

\subsection{Rayleigh-Taylor instability of inverted film}
\label{sec:rt}

The instabilities we have discussed so far are discussed in the fluid configurations where contact line is present, 
and an obvious question is what happens if there is no contact line, that is, if we have an infinite film spreading 
down an (inverted) surface. This configuration is expected to be susceptible to Rayleigh-Taylor (R-T) type of instability 
since we effectively have a heavier fluid (liquid) above the lighter one (air).   The question is what are the 
properties of this instability for a film flowing down an inverted surface, and how this instability relates to the 
contact line induced one, discussed so far.

\begin{figure}[htb]
\centering
\includegraphics[scale=0.4, angle=-90]{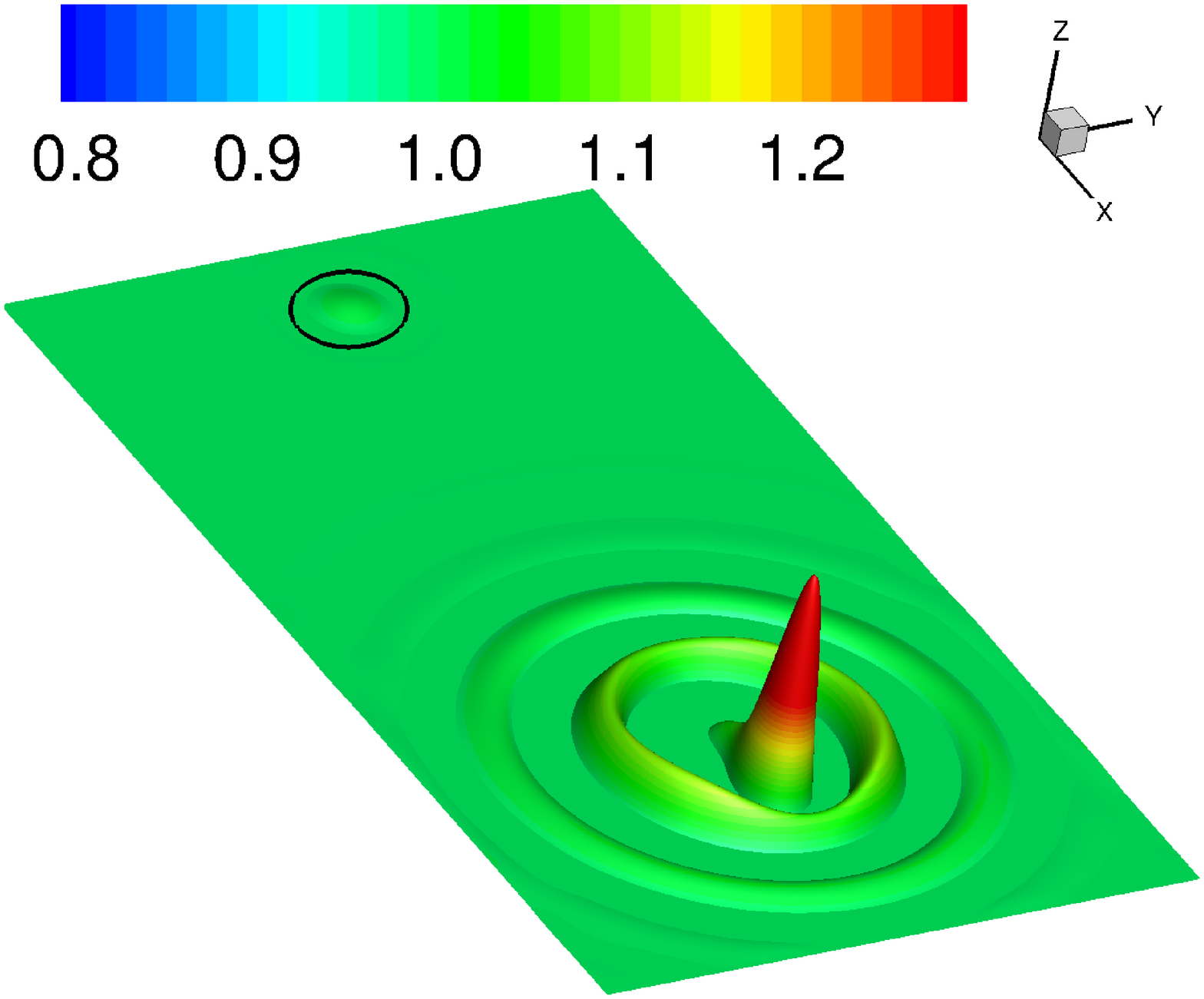}
\caption{(Color online) Simulation of Rayleigh-Taylor instability for hanging film on inclined 
plane at $D=-1.0$. The black circle indicates the initial profile. The surface shown in 
the downstream is the surface profile at $t=20$. The computational domain is 
$[0,90] \times [0,50]$.}
\label{fig:RT}
\end{figure}

Figure~\ref{fig:RT} shows the results of 3D simulation of an infinite film (no contact line). The initial condition 
at $t=0$ is chosen as a flat film with localized half-sphere-like perturbation of amplitude $0.1$  (marked by the 
black circle in Fig.~\ref{fig:RT}). As an example, we use  $D=-1.0$. At time $t=20$, the perturbation had been 
amplified, as expected. The properties of this instability are, however, very different from the one observed in 
thin film flow where contact line is present. For this $D$, if a contact line is present, we see only a capillary 
wave behind the front (vis. Fig.~\ref{fig:HF1} and Fig.~\ref{fig:HF4}), and we do not find upstream propagating waves 
as for the infinite film shown in Fig.~\ref{fig:RT}. Therefore, the instability discussed so far is not of R-T type - 
instead, it is induced by the presence of a contact line.   

One obvious question is why we do not observe (additional) R-T instability in the flow with fronts. The answer is 
that the speed with which a perturbation on a main body of a film (such as infinite film shown in Fig.~\ref{fig:RT}) 
propagates downstream is faster than the speed of the contact line itself. Therefore, in the simulations of films 
with fronts presented so far, flat film instabilities do not have time to develop since they reach the contact 
line before having a chance to grow. We note that in Fig.~\ref{fig:RT} we used large scale perturbation to illustrate 
the point; in a physical problem, one would expect surface perturbations to be characterized by much smaller amplitudes 
and would therefore require much longer time to grow to the scale comparable to the film thickness. As a consequence, 
R-T instability could be expected to become relevant only for the films characterized by the spatial extend which is 
much larger than the one considered here.    Similar conclusion extends to stability of infinite rivulets, discussed 
briefly in Sec.~\ref{sec:riv}.

\section{\label{sec:noniso}Inverted film of variable viscosity with a front}

\begin{figure}[htb]
\centering
\includegraphics[scale=0.4]{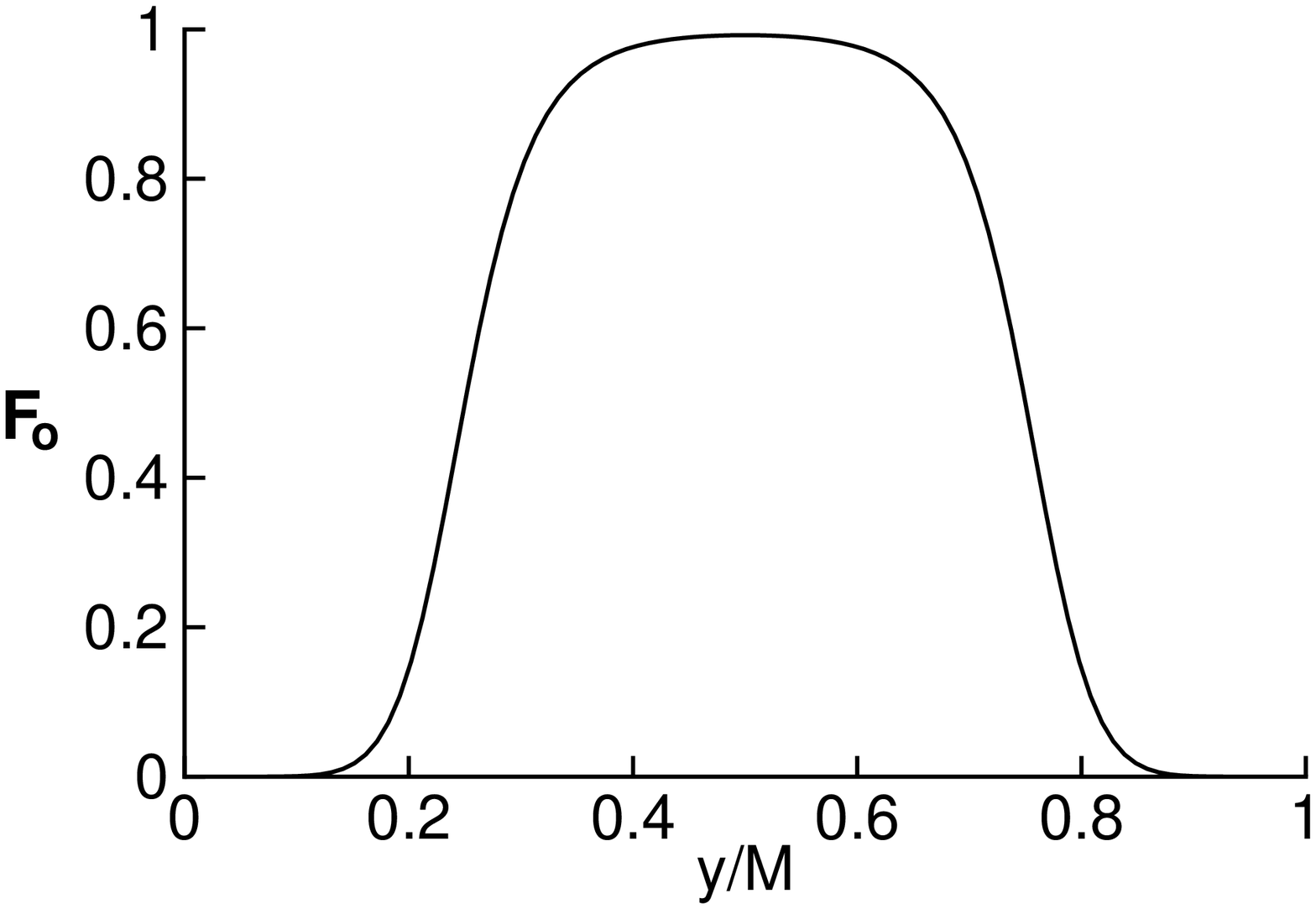}
\caption{Entrance profile of film thickness.}
\label{fig:entrance}
\end{figure}

Consider now a situation, when the film is of finite width and the fluid viscosity is variable in the 
transverse direction, ie., $\bar{\mu}=\bar{\mu}(y)$ in Eq.~(\ref{eq:thin_film_3d}). We have to substitute 
the no-flow boundary conditions, Eq.~(\ref{eq:full3d_by}), at the borders of the computational domain with 
the following ones
\begin{eqnarray*}
& h(x,0,t)=h(0,M,t)=b,& \\ 
& h_x(x,0,t)=h_x(0,M,t)=0,&
\end{eqnarray*}
The flow starts at the top of the domain according to the condition
\begin{equation}
 h(0,y,t)=b + F_0\left(\frac{y}{M}\right)\frac{(at)^2}{1+(at)^2},
\label{eq:entrance}
\end{equation}
where the value of parameter $a$ is $0.775$ and the bell shape of the entrance thickness profile determined by 
the function $F_0$ is shown in Fig.~\ref{fig:entrance}. 
Equation~(\ref{eq:entrance}) mimics the growth of the flow rate and the film thickness at the entrance boundary as the 
liquid starts being delivered to the substrate during the ramp-up in industrial processes or experiments. Characteristic 
patterns of flow depend on the value of parameter $D$, width of the film and distribution of viscosity, and establish when 
the product $at$ in Eq.~(\ref{eq:entrance}) reaches values of several units.

In many practical situations, the variation of viscosity is a result of its dependence on temperature, which is 
usually non-linear. For example, assume that the film temperature $T$ drops linearly with the lateral coordinate $y$ 
and the viscosity (in $Pa\cdot s$) is given by a generic equation:
\begin{eqnarray}
 T &=& T_0 - ay, \quad a=\mbox{const.}; \label{eq:t1}\\
 \log_{10}(\mu) &=& -A+\frac{B}{T-C},
\end{eqnarray}
where $A=5.5$, $B=700$ K and $C=52$ K are the material constants, and $T$ is the liquid temperature. The viscosity function 
is normalized by its value at $y=0$. The constant value $T_0$ is assumed to be equal $153$ K, while the lateral temperature 
drop across the computational domain is assumed to be equal $11$ K in all cases discussed hereafter, resulting in $7.5$ fold 
viscosity growth from bottom to top of the domain. 

\begin{figure}[htb]
\centering
\subfigure[$t=9.4$.]{
\includegraphics[scale=0.35, angle=-90]{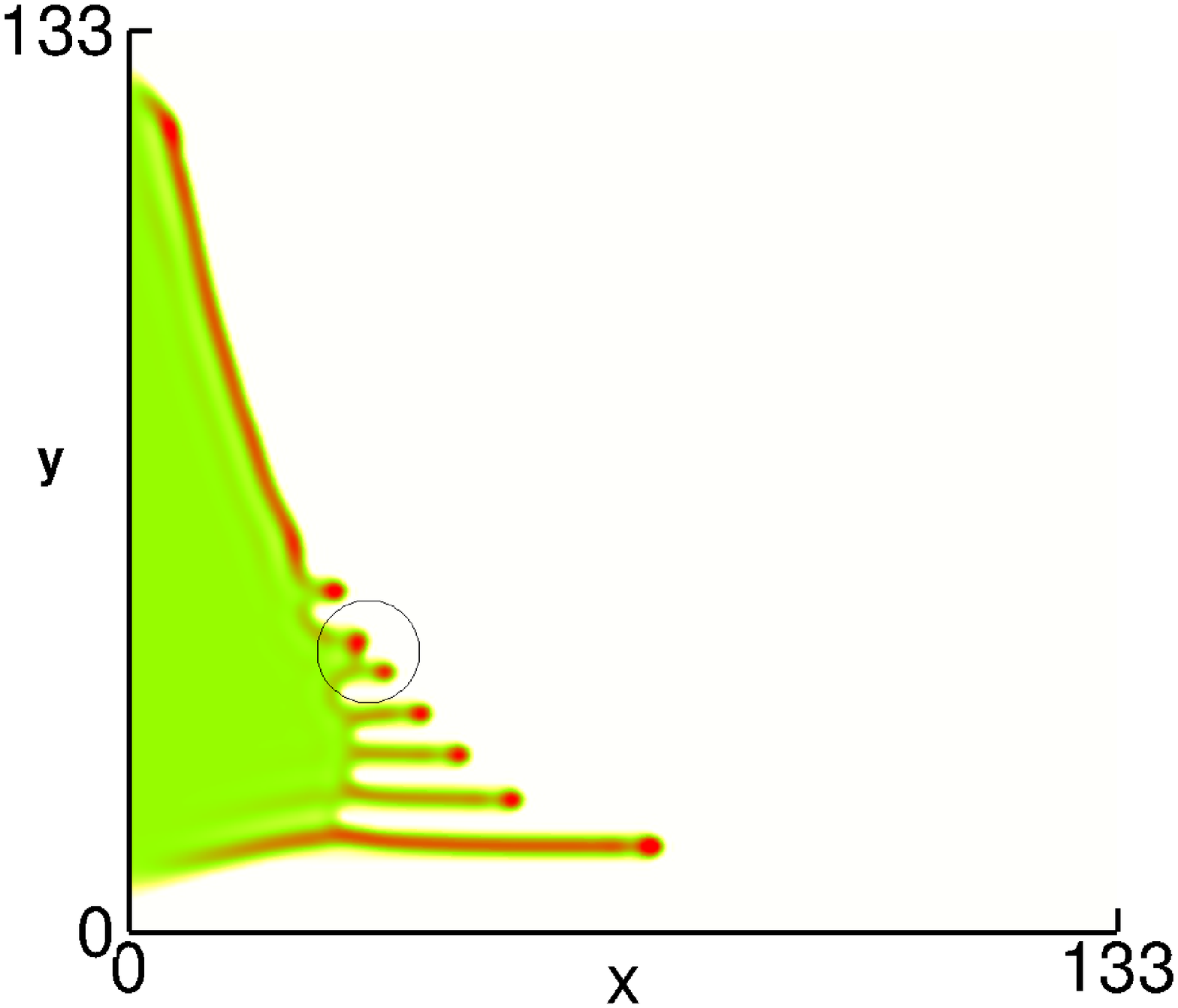}}\\
\subfigure[$t=16.5$.]{
\includegraphics[scale=0.35, angle=-90]{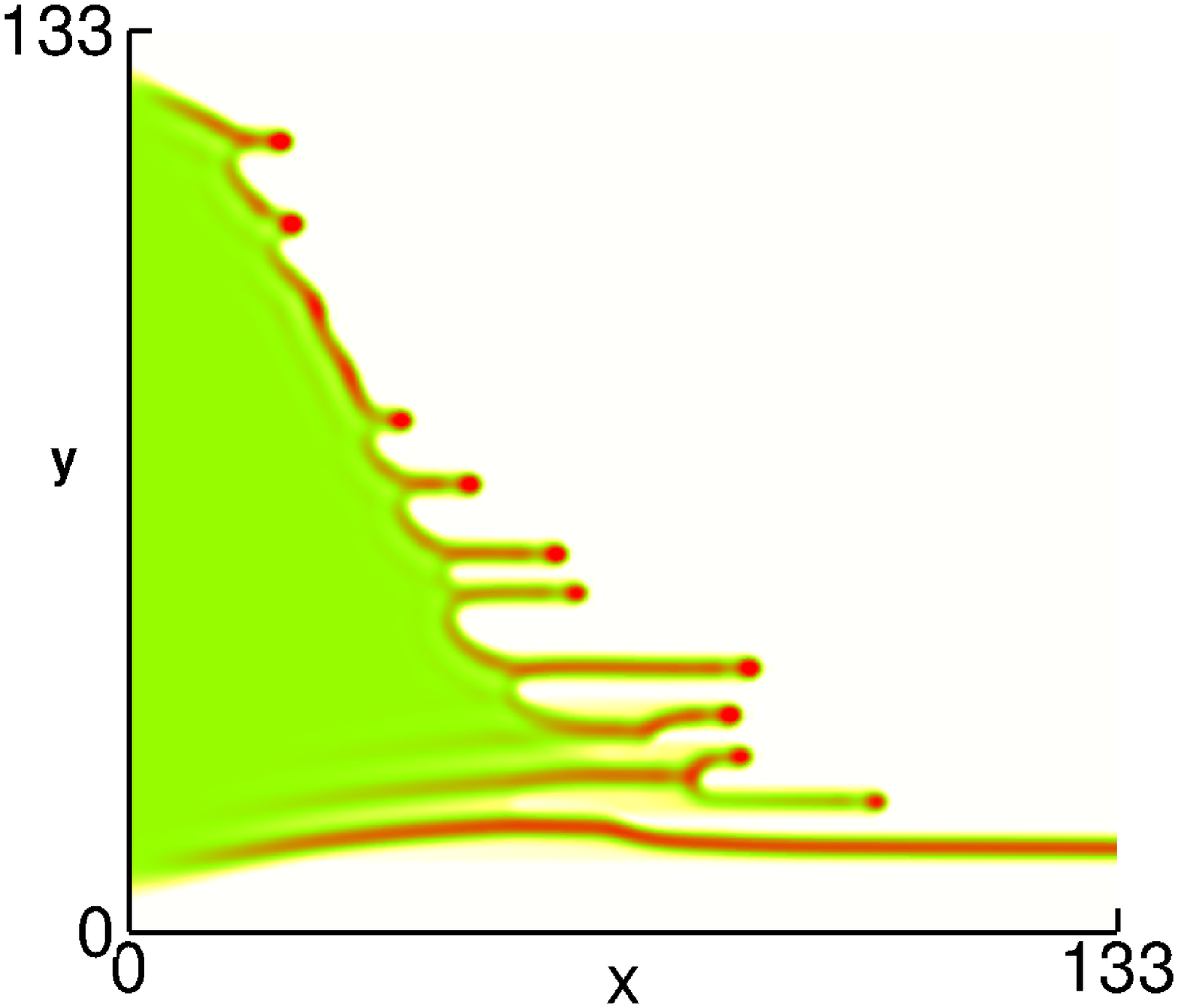}}
\caption{(Color online) Film thickness for  $D=-0.88$.  The circle in the part (a) indicates two coalescing fingers.}
\label{fig:iso_088}
\end{figure}

Similarly to isoviscous cases discussed earlier, the film front is always unstable, producing finger-like 
rivulets. The morphology of the non-isoviscous film (as compared to an isoviscous case) stems from the fact 
that similar structures such as fingers at different parts of the film move with different speeds. As an example, 
Fig.~\ref{fig:iso_088} (a) and (b) show distribution of film thickness at $t=9.4$, $16.5$, with parameter $D=-0.88$ and the 
dimensionless domain width $M$ equal to $133$. Morphology of individual fingers is similar to the one of 
fingers in isoviscous cases shown in Fig.~\ref{fig:HF4}, but there is obvious mass redistribution along the front 
line because of its inclination, resulting in coalescence of some of the fingers and variation of the 
finger-to-finger distance. A finger produced by coalescence of two parent fingers, such as finger $5$ in 
Fig.~\ref{fig:iso_088}(b), has a higher flow rate and moves faster than its neighbors.

\begin{figure}[htb]
\centering
\includegraphics[scale=0.35, angle=-90]{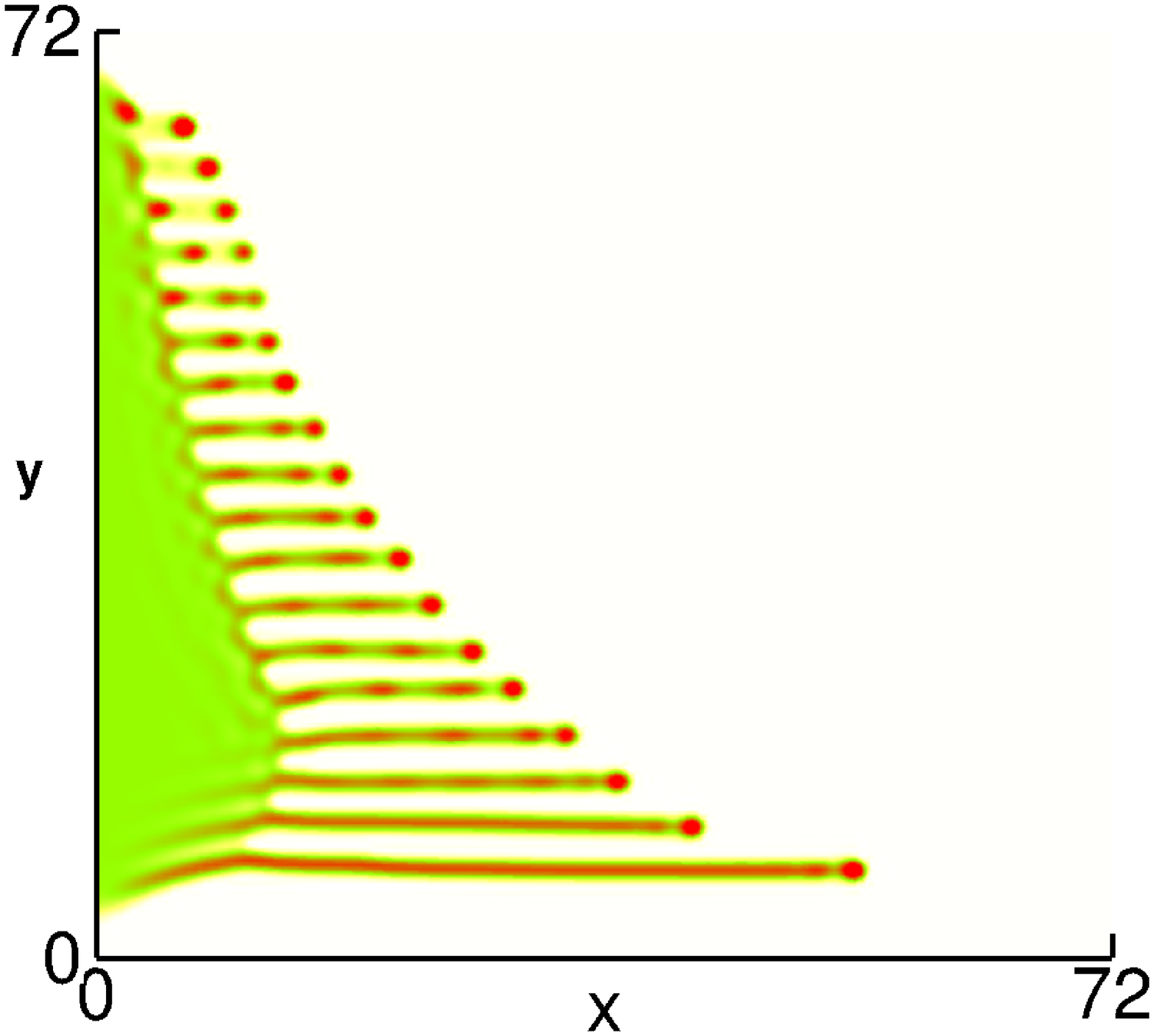}
\caption{(Color online) Film thickness for $D=-1.5$ at $t=25.4$.}
\label{fig:iso_15}
\end{figure}

These specific properties of non-isoviscous film are common for flows in the whole spectrum of parameter $D$, but morphology 
of individual fingers strongly depends on the type of the flow, similar to already considered isoviscous cases. 
Figure~\ref{fig:iso_15} shows distribution of film thickness for parameter $D=-1.5$, which corresponding to {\it type 2}, 
with film width $M=72.2$, at $t=25.4$. The leading head capillary ridges are followed by a rivulet with followup smaller waves 
moving faster than the leader. The pattern is common for all fingers, but the speed of propagation increases with temperature. 
Each of the fingers has a faster-moving neighbor on the higher temperature side, causing slight increase of the background film 
thickness on that side compared to the colder side. As a result, the fingers may have a tendency of being diverted and coalesce 
with warmer neighbors resulting in mass transport from colder areas to warmer areas of the flow, despite the fact that the film 
velocity does not explicitly depend on the temperature gradient.

\begin{figure}[htb]
\centering
\includegraphics[scale=0.35, angle=-90]{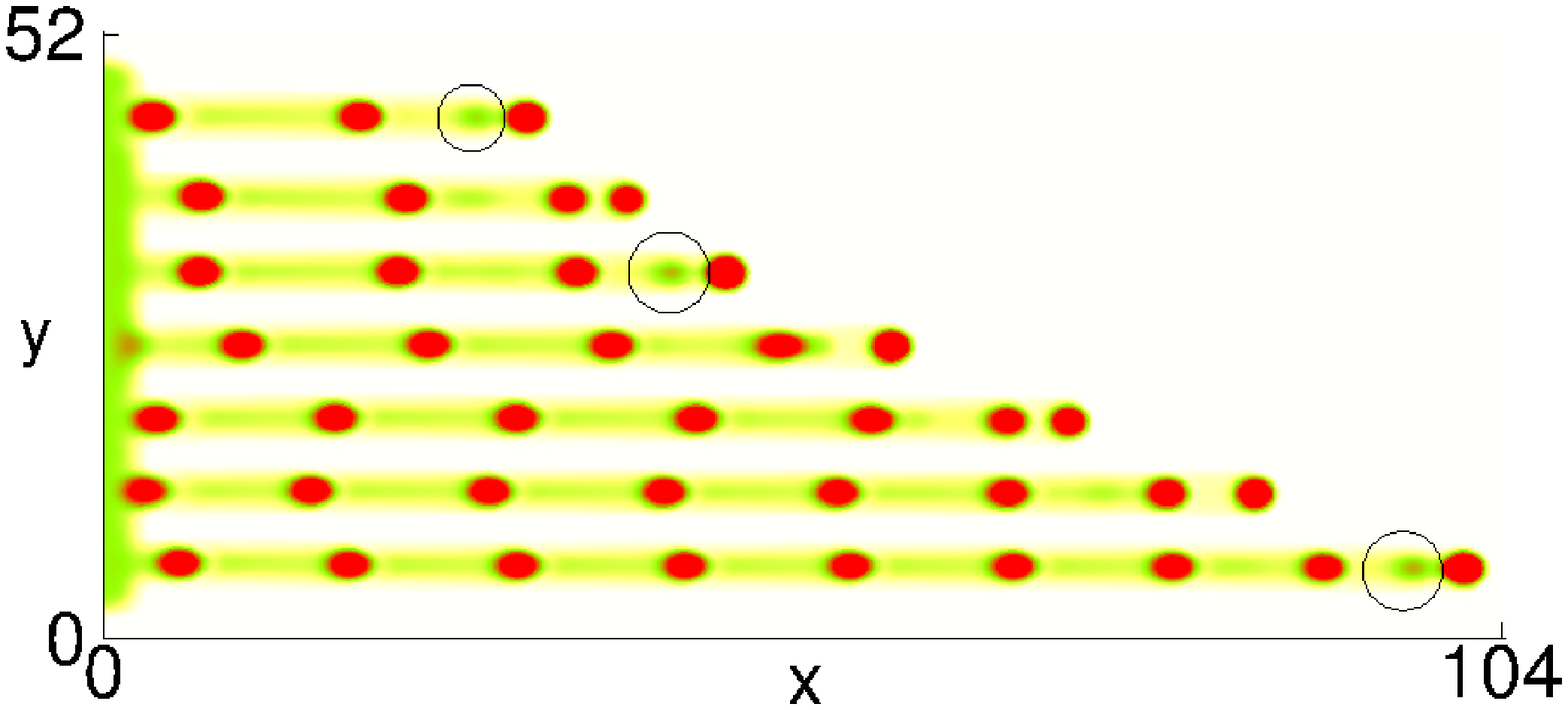}
\caption{(Color online) Film thickness for $D=-2.54$ at $t=44.4$.}
\label{fig:iso_254}
\end{figure}

In {\it type 3} film flow, the height of follow-up drops is already close to that of finger head drops, as shown 
in Fig.~\ref{fig:iso_254}, showing thickness distribution in film with parameter $D=-2.54$, film width $M=52.2$ at 
$t=44.4$. In {\it type 3} flows, there is no propagation front and the flow itself consists of a series of 
propagating fingers. Another peculiarity of this type of flows is existence of small drops, separating from the 
leading drops to be immediately consumed by the following drop in the train, as indicated by circles in 
Fig.~\ref{fig:iso_254}. These features are also observed in isoviscous flows for similar values of $D$; see, e.g., 
cross sectional profile in Fig.~\ref{fig:HF3}. The viscosity effect in the developed {\it type 3} flows shows mainly as 
the difference in speeds for droplets in low and high viscosity regions.

\section{\label{sec:conclusion}Conclusions}

In the previous work~\cite{lin_pof10} we carried out extensive computational and asymptotic analysis of the two dimensional 
flow of a completely wetting fluid down an inverted surface. Complex behavior was uncovered with different families of waves 
evolving in the configurations characterizing by different values of the governing parameter $D$. In the present work, we 
have considered fully three dimensional problem of spreading down an inverted surface. We find that there is an elaborate 
interaction of surface instabilities and contact line instabilities. For the values of $D$ which are not too small 
(approximately $D \ge -1.1$) we find similar behavior as already known for the flow down an inclined surface, with the main 
difference that the finger-like patterns that form are spaced
more closely and the fingers themselves are more narrow for negative $D$'s.   As $D$ is further decreased, we
still find instabilities of the contact line leading to formation of fingers, but in addition we observe formation of 
surface waves, which propagate down the fingers with the speed larger than the speed of the fingers themselves:
therefore, these propagation waves (which may appear as drops on top of the base film) travel down a finger, reach
the front and merge with the leading capillary ridge.   Behind the fingers, in this regime we find strip-like waves 
(whose fronts are independent of the transverse direction).   These waves are convective in nature and leave 
behind a portion of a flat film whose length increases with time.    For even smaller $D$'s (smaller than approximately
$-2.0$), these transverse strip-like waves disappear, and the whole  film is covered by localized waves.
These localized waves travel faster than the film itself, and converge towards the fingers which form at the front.

It is worth emphasizing that the properties of the surface waves which form due to the presence of fronts are different
from the ones which would be expected if the fronts were not present.   To illustrate this effect, we consider an infinite
film with a localized perturbation which is expected to be unstable by a Rayleigh-Taylor type of instability.   We find that
this instability leads to a different type of surface waves, which may or may not be observable in physical experiments, 
depending on the size of the fluid domain.   

In the second part of the paper we consider flow where fluid viscosity is not constant, but varies in the transverse direction.
The most important difference is the loss of flow periodicity in the lateral direction. The fingers 
in the warmer parts of the flow move faster than those in the colder areas, yielding slight increase of the background film 
thickness on warmer side of each finger compared to the colder side. This results in mass redistribution from colder areas to 
warmer areas of the flow, more pronounced for lower values of $|D|$, despite explicit independence of the film velocity on the 
temperature gradient.

{\it Acknowledgments.}
This work was partially supported by NSF grant No. DMS-0908158.

\appendix
\section{\label{sec:ADI} Numerical method for 3D thin film equation}

\subsection{Solving nonlinear time dependent PDE}
In general, time dependent PDE, Eq.~(\ref{eq:thin_film_3d}), can be expressed as $h_t + f(h) = 0$, where 
$h=h(\vect{x},t)$ is the unknown function with time variable $t$, spatial variables $\vect{x}$, and $f$ 
is a nonlinear discretization operator for spatial variables. From time $t^n=n\Delta t$ to $t^{n+1}$, where 
$\Delta t$ is the time step, the PDE can be integrated numerically by the so-called $\theta$ method leading 
to a nonlinear system:
\begin{equation}
h^{n+1} + (1-\theta) \Delta t f^{n+1} = h^n - \theta\Delta t f^n,
\label{nonlinear_equation}
\end{equation}
where $f^n=f(h^n)$ and $h^n=h(\vect{x}, t^n)$. To solve the nonlinear system 
(\ref{nonlinear_equation}), we apply the Newton's method. Firstly, we linearize $h^{n+1}$ about a guess for the 
solution by assuming $h^{n+1} = h^* + c$, where $h^*$ is a guess and $c$ is the correction. Then we express 
the nonlinear part using Taylor's expansion
\[
f^{n+1}=f(h^*+c) \approx f(h^*)+J_f(h^*)\cdot c = f^*+J_f^* \cdot c,
\]
where $J_f$ is the Jacobian matrix for function $f$ and $J_f^*=J_f(h^*)$. After substituting the above quantities 
into Eq.~(\ref{nonlinear_equation}), we obtain a linear system for the correction term, $c$:
\begin{equation}
\lp I + (1-\theta) \Delta t J_f^* \rp c = -h^* - (1-\theta)\Delta t f^* + h^n 
- \theta\Delta t f^n, 
\label{nonlinear_iteration}
\end{equation}
$I$ is the identity matrix that has the same size as the Jacobian matrix, $J^*_f$. The solution at $t=t^{n+1}$ is 
obtained by correcting the guess iteratively until the process converges, ie., the new correction is small enough. 

\subsection{Spatial discretization}
We discretize the spatial derivatives of thin film equation, Eq.~(\ref{eq:thin_film_3d}), through finite difference 
method. The grid points in the computational domain, $[0,L]\times [0,M]$, is defined as 
\beas
 x_i &=& \left(i-\frac{1}{2}\right)\Delta x, \quad i=1,\cdots,n_x,\\
 y_j &=& \left(j-\frac{1}{2}\right)\Delta y, \quad j=1,\cdots,n_y,
\eeas
where $\Delta x=L/n_x$, $\Delta y=M/n_y$ are the step size in the $x$, $y$ domain; $n_x$, $n_y$ are 
number of grid points in the $x$, $y$ domain, respectively. 

The scheme presented here is 2nd order central difference scheme. In the following, the subscripts, $i,j$, denote that 
the value been taken at $(x_i,y_j)$. The notation $h_{i+1/2,j}$ denotes an average at the point $(x_{i+1/2},y_j)$ as
\[
 h_{i+1/2,j} = \frac{h_{i+1,j}+h_{i,j}}{2}.
\]
Also we use the standard difference notation as
\beas
 \delta_xh_{i+1/2,j} &=& h_{i+1,j}-h_{i,j},\\
 \delta_yh_{i,j+1/2} &=& h_{i,j+1}-h_{i,j},\\
 \delta^3_xh_{i+1/2,j} &=& h_{i+2,j}-3h_{i+1,j}+3h_{i,j}-h_{i-1,j},\\
 \delta^3_yh_{i,j+1/2} &=& h_{i,j+2}-3h_{i,j+1}+3h_{i,j}-h_{i,j-1},\\
 \delta_x\delta^2_yh_{i+1/2,j} &=& h_{i+1,j+1}-2h_{i+1,j}+h_{i+1,j-1}\\
 & &-h_{i,j+1}+2h_{i,j}-h_{i,j-1},\\
 \delta_y\delta^2_xh_{i,j+1/2} &=& h_{i+1,j+1}-2h_{i,j+1}+h_{i-1,j+1}\\
 & &-h_{i+1,j}+2h_{i,j}-h_{i-1,j}.
\eeas

The discretization of each term in Eq.~(\ref{eq:thin_film_3d}) involving spatial derivatives is as follows:\\
The surface tension term
\beas
 & \nabla\cdot[h^3\,\nabla\nabla^2 h]_{i,j}= &\\
 & \left(h^3_{i+1/2,j}\delta^3_xh_{i+1/2,j}-h^3_{i-1/2,j}\delta^3_xh_{i-1/2,j}\right)/\Delta x^4 &\\
 & \left(h^3_{i+1/2,j}\delta_x\delta^2_yh_{i+1/2,j}-h^3_{i-1/2,j}\delta_x\delta^2_yh_{i-1/2,j}\right)/\Delta x^2\Delta y^2 &\\
 & \left(h^3_{i,j+1/2}\delta_y\delta^2_xh_{i,j+1/2}-h^3_{i,j-1/2}\delta_y\delta^2_xh_{i,j-1/2}\right)/\Delta x^2\Delta y^2 &\\
 & \left(h^3_{i,j+1/2}\delta^3_yh_{i,j+1/2}-h^3_{i,j-1/2}\delta^3_yh_{i,j-1/2}\right)/\Delta y^4. &
\eeas
The normal gravity term
\beas
 &\nabla\cdot[h^3\,\nabla h]_{i,j}= &\\
 & \left(h^3_{i+1/2,j}\delta_xh_{i+1/2,j}-h^3_{i-1/2,j}\delta_xh_{i-1/2,j}\right)/\Delta x^2 &\\
 & + \left(h^3_{i,j+1/2}(h_{i,j+1}-h_{i,j})-h^3_{i,j-1/2}(h_{i,j}-h_{i,j-1})\right)/\Delta y^2.&
\eeas
The tangential gravity term
\begin{equation}
 \frac{\partial}{\partial x}\left(h^3\right)_{i,j} = \left(h^3_{i+1/2,j}-h^3_{i-1/2,j}\right)/\Delta x.
\end{equation}

\subsection{Fully implicit algorithm}

Applying algorithm Eq.~(\ref{nonlinear_iteration}) to 3D thin film Eq.~(\ref{eq:thin_film_3d}), we have~\cite{DK_jcp02}
\bea
\lp I + (1-\theta) \Delta t (J_{fx}^*+J_{fy}^*+J_{fm}^*) \rp \cdot c = \nonumber\\
-h^* - (1-\theta)\Delta t f^* + h^n - \theta\Delta t f^n,
\label{3d_newton}
\eea
where $J_{fx}$, $J_{fy}$ and $J_{fm}$ are the Jacobian matrices for $x$, $y$ and mixed derivative terms of 
function $f$, respectively. Equation~(\ref{3d_newton}) is a non-symmetric sparse linear system that has 
$n_xn_y$ unknowns. For large $n_x$ and $n_y$, as it is the case for the problems discussed in the present work, 
solving such a system carries a significant computational cost. In general, the operation count is proportional 
to $O(n_x^3n_y^3)$ or $O(n_x^2n_y^2)$, depending on the matrix solver.  

\subsection{ADI method}
To decrease the computational cost, Witelski and Bowen suggested to use the approximate-Newton 
approach~\cite{witelski_anm03}. The idea is to replace the Jacobian matrix by an approximated one
\bea
 & \left[I+(1-\theta)\Delta t\left(J^*_{fx}+J_{fy}^*+J_{fm}^*\right)\right] & \nonumber\\
 & \sim \left[I+(1-\theta)\Delta t J^*_{fy}\right] \left[I+(1-\theta)\Delta t J^*_{fx} \right] &.
\eea
Therefore we get a new system of equations
\be
\left[I+(1-\theta)\Delta t J^*_{fy}\right] \left[I+(1-\theta)\Delta t J^*_{fx}\right]
\cdot c = R,
\label{3d_adi}
\ee
where $R=-h^*-(1-\theta)\Delta tf^*+h^n-\theta\Delta tf^n$ is the right hand side of Eq.~(\ref{3d_newton}). 
One should note that as long as $c$ decreases after each iteration and approaches $0$ in some norm, we have $R=0$,
leading to Eq.~(\ref{nonlinear_equation}). That is, such an approach does not affect the stability and 
accuracy of the original space-time discretization. 

Under the same spirit of alternating direction implicit method (ADI), equation~(\ref{3d_adi}) can be 
easily split into two steps:
\bea
\left(I+(1-\theta)\Delta t J_{fx}^*\right) \cdot w &=& R,\nonumber\\
\left(I+(1-\theta)\Delta t J_{fy}^*\right) \cdot c &=& w.
\label{eq_ADI}
\eea
The main advantage of such splitting is that the operations in the $x$ and $y$ directions are decoupled and 
therefore the computational cost reduced significantly. Specifically for our discretization, the Jacobian matrices 
in the $x$ and $y$ direction are penta-diagonal matrices, leading to system that can be solved in $O(n_x)$ and  
$O(n_y)$ arithmetic, and the overall computational cost for solving Eq.~(\ref{eq_ADI}) is proportional to 
$O(n_xn_y)$. 

The approach presented here deals with a matrix that is an approximation to the original Jacobian one. Therefore 
we should not expect the convergent rate of the ADI method to be quadratic. However, since the approximation error 
is proportional to $O(\Delta t)$, for small enough time step, we expect the rate of convergence to be close to 
quadratic; see~\cite{witelski_anm03} for further discussion of this issue. Furthermore, the ratio in the operation 
count between fully implicit discretization and the ADI method is $O(n_xn_y)$. That is, even if we need to decrease 
the time step or increase the number of iterations to achieve convergence, the ADI method is still more efficient as 
long as the additional effort is of $o(n_xn_y)$. In our experience, under the same conditions, ADI method is 
significantly more efficient compared to fully implicit discretization.

\bibliographystyle{unsrt}
\bibliography{films}

\end{document}